# Measurement of conditional phase shifts for quantum logic

Q. A. Turchette*, C. J. Hood, W. Lange, H. Mabuchi, and H. J. Kimble
*Norman Bridge Laboratory of Physics 12-33*
*California Institute of Technology, Pasadena, CA 91125*
(October 27, 1995)

Measurements of the birefringence of a single atom strongly coupled to a high-finesse optical resonator are reported, with nonlinear phase shifts observed for intracavity photon number much less than one. A proposal to utilize the measured conditional phase shifts for implementing quantum logic via a quantum-phase gate (QPG) is considered. Within the context of a simple model for the field transformation, the parameters of the "truth table" for the QPG are determined.

Although the theory of quantum computation dates back more than a decade to the seminal works of Feynman and Deutsch [1], there has recently been an explosion of new activity driven in large measure by Shor's quantum algorithm [2] for efficient factorization. While most attention has been directed toward theoretical issues, several strategies have also been proposed for laboratory investigations [3]. However, the demands on experimental systems for building quantum computational networks [4] are quite severe, requiring strong coupling between quantum carriers of information ("qubits") in an environment with minimal dissipation. Hence, experimental progress has lagged behind the remarkable theoretical developments in quantum information theory.

Within this context, we present a significant experimental step toward realizing quantum logic with individual photons as qubits. Moreover, our work bears import for related experimental challenges such as quantum nondemolition (QND) measurement and quantum cryptography. Specifically, we report the demonstration of conditional dynamics *at the single photon level* between two frequency-distinct fields in an optical resonator. Our measurements utilize the circular birefringence of an atom strongly coupled to the resonator to rotate the linear polarization of a transmitted probe beam. The phase shift between circular polarization states $\sigma_\pm$ is conditioned upon the intensity of a pump beam via a Kerr-type nonlinearity, with conditional phase shifts $\Delta \sim 16°$ per intracavity photon extracted from our data. To explore further the prospects for quantum logic based on these capabilities, we have experimentally investigated a candidate quantum-phase gate (QPG) and, within the context of a simple model, have extracted relevant phase shifts for the "truth table" of the QPG. In our proposed implementation, "flying qubits" are single-photon pulses propagating in two frequency-offset channels, with internal states specified by $\sigma_\pm$ polarization.

It should be noted at the outset that necessary and sufficient testing procedures have not yet been established for providing *direct experimental verification* that a given "black box" laboratory system can perform quantum logic transformations with sufficient fidelity to implement Deutsch's Quantum Turing Machine [1]. In particular, it is not known what level of dissipation (if any) can be tolerated in experimental systems before the advantages of unitary information processing are lost. However, any laboratory quantum gate must exhibit coherence and demonstrably produce entanglement between qubits. The practical application of such criteria requires the formulation of new measurement strategies, which we consider explicitly for our experiment.

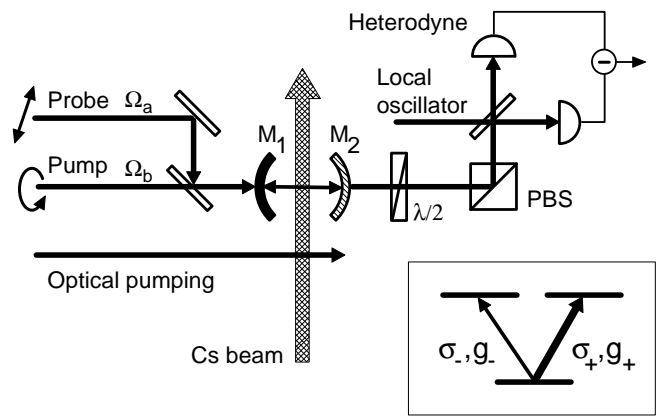

FIG. 1. Schematic of the experimental apparatus.

Our efforts here focus on the implementation of quantum logic by exploiting the extremely large optical nonlinearities realizable in cavity quantum electrodynamics (CQED) [5,6]. In CQED systems, individual photons circulating in a high-finesse resonator can interact strongly via their mutual coupling to a single intracavity atom. The critical parameters which characterize our apparatus are $g$, the dipole coupling rate of atom to cavity; $\kappa$, the cavity-field damping rate; and $\gamma$, the transverse atomic decay rate to non-cavity modes. The current work is performed with parameters such that $\kappa > g^2/\kappa > \gamma$. In this *bad cavity regime* the atom's coherent coupling to the cavity mode (at rate $g^2/\kappa$) dominates incoherent emission into free-space (at rate $\gamma$), making it possible to couple strongly a single atom to the cavity mode in a manner which allows for efficient transfer of electromagnetic fields from input to output channels (at rate $\kappa$), thus creating an effectively *one-dimensional atom* [6]. The atom-cavity system may therefore be viewed as a quantum-optical de-vice (a *nonlinear one-atom waveplate*) which is exploited for processing field states.

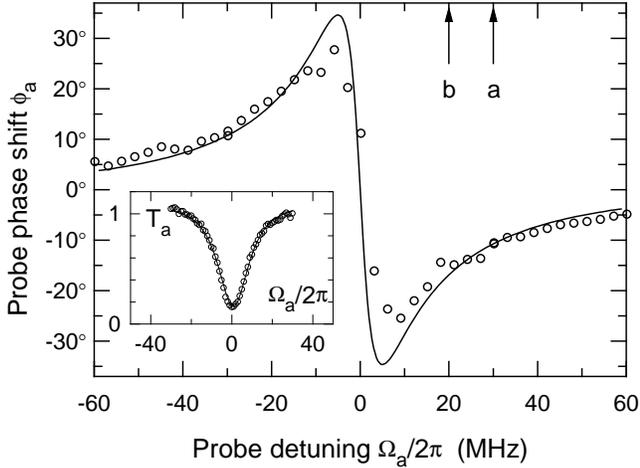
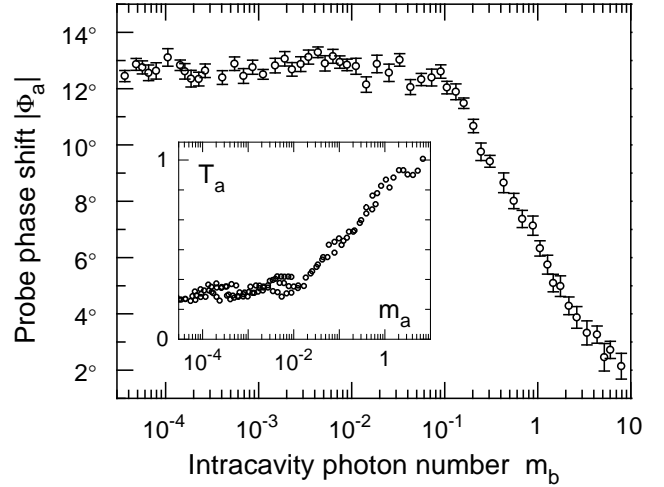

FIG. 2. Measured weak-field response of the atom-cavity system for $\overline{N} = 1.0 \pm 0.1$ atoms. Full curves represent the theoretical model from Ref. [6]. The inset shows the squared modulus of the normalized probe transmission $T_a$ and the main axes show probe phase shift $\phi_a$. $\Omega_a$ denotes detuning from the resonance frequency $\omega_A = \omega_C$.

FIG. 3. Probe phase shift $\Phi_a$ vs. $m_b$ for an injected $\sigma_+$ pump, for $\overline{N} = 0.9$ atoms and pump (probe) detuning of $+20$ ($+30$) MHz from atomic resonance as shown in Fig. 2. Error bars indicate uncertainties in least squares fits used for evaluating the phase shifts. The inset shows transmission $T_a$ vs. $m_a$ for a resonant probe without pump, with $\overline{N} = 0.6$.

Conditional dynamics in our system originate from the nonlinear optical response of a Cesium atom coupled to the cavity field. For the particular optical frequencies used, the relevant atomic states form a three-level system shown in the inset to Fig. 1. The transitions couple to cavity modes with orthogonal circular polarizations $\sigma_\pm$ with rates $g_\pm$, where the $\sigma_+$ transition corresponds to $(6S_{1/2}, F = 4, m = 4) \to (6P_{3/2}, F' = 5, m' = 5)$ and the $\sigma_-$ transition connects $(m = 4)$ to $(m' = 3)$. Since $g_- = g_+/\sqrt{45}$, we set $g_- = 0$ to simplify the current discussion (this is not an essential approximation). As shown in Fig. 1, the ground state $(F = 4, m = 4)$ is prepared by optical pumping of an atomic beam of Cesium just before it enters the high finesse ($\mathcal{F} = 18,000$) cavity. The cavity length and Gaussian waist are 56 $\mu$m and 35 $\mu$m. The mirrors ($M_1, M_2$) have transmission coefficients $(1.1 \times 10^{-6}, 3.5 \times 10^{-4})$. Together with the atomic lifetime $\tau = 32$ ns and transit time $T_0 = 7\tau$, these parameters lead to the set of rates $(g_+, \kappa, \gamma, T_0^{-1})/2\pi = (20, 75, 2.5, 0.7)$ MHz. Hence, the intracavity saturation photon number $m_0 = 4\gamma^2/3g_+^2 = 0.02$ photons, the critical atom number $N_0 = 2\kappa\gamma/g_+^2 = 0.94$ atoms, and the one-photon tipping angle $2g_+T_0 = 15\pi$.

To characterize photon-photon interactions inside our atom-cavity device, we investigate the transmission of monochromatic coherent-state pump and probe beams which are independently tunable in frequency, power and polarization [6] (see Fig. 1). After passing through the cavity, these beams are analyzed for polarization state with a rotatable half-wave plate, a polarizer, and balanced heterodyne detectors.

Turning now to our measurements, we present in Fig. 2 the weak-field response (average intracavity photon number $\ll m_0$) of the atom-cavity system for the case of coincident atomic ($\omega_A$) and cavity ($\omega_C$) resonances. For these scans the average intracavity atom number is $\overline{N} = 1.0 \pm 0.1$ atoms, as determined by fits to the data as discussed in Refs. [6,7]. The inset data in Fig. 2 give the ratio $T_a$ of transmitted power with atoms present to that without as a function of the detuning $\Omega_a$ of the probe, which is $\sigma_+$-polarized to interact with the strong $g_+$ transition. The main data of Fig. 2 represent the phase of the transmission function and are taken by injecting a linearly-polarized probe beam, with the $\sigma_+$ component of this beam attaining a phase shift due to the composite atom-cavity system, while the $\sigma_-$ component only receives a phase shift corresponding to an empty cavity (in the approximation $g_- \to 0$). The differential phase $\phi_a$ between the $\sigma_\pm$ components combines with changes in amplitude to produce an elliptically-polarized output beam with its major axis rotated by $\phi_a/2$ relative to the linearly-polarized input, so that $\phi_a$ can be determined by analysis of the polarization state of the output beam.

To utilize these phase shifts for conditional dynamics, we next consider measurements of nonlinear dispersion. We fix the detuning $\Omega_a$ of the weak linearly-polarized probe beam ($m_a \approx 10^{-4}$ photons) at a position on the dispersion curve of Fig. 2 corresponding to relatively low intracavity loss as determined from $T_a$. As a controlling field, we inject a $\sigma_+$-polarized pump beam at detuning $\Omega_b$. Figure 3 displays the variation of the phase $\Phi_a$ of the probe beam for a wide range of pump powers, with $\Phi_a$ measured by polarization interferometry as discussed above. In the limit $m_b \to 0$, $\Phi_a \to \phi_a$ which is the phase shift for the probe field alone. Note that the pump-probe coupling is manifest for $m_b \approx 0.1$, with a 30% reduction of $|\Phi_a|$ as $m_b$ goes from 0.1 to 0.3 photons. The Fig. 3 inset shows the corresponding nonlinearity of $T_a$ for single frequency resonant excitation.

These measurements represent the realization of a nonlinear optical susceptibility at the single photon level and unambiguously demonstrate the conditional dynamics necessary for implementing quantum logic. To quantify further the interaction strength involved, we note that the pump and probe input fields are prepared as uncorrelated coherent states with small amplitudes $|\alpha|^2, |\beta|^2 \ll 1$. Hence, their composite state can be expanded in the form $|\psi\rangle \propto [|0\rangle_a + \alpha|1\rangle_a] \otimes [|0\rangle_b + \beta|1\rangle_b]$. Our *ansatz* for the transformation of field states is

$$|j\rangle_a|k\rangle_b \mapsto e^{i\mu_{jk}}|j\rangle_a|k\rangle_b, \qquad j,k = \{0,1\}, \quad (1)$$

which amounts to the physically-motivated assumption that Fock states asymptotically connect to the dressed states of the atom-cavity system and hence are the appropriate eigenstates of the transformation. For $\mu_{00} + \mu_{11} \neq \mu_{01} + \mu_{10}$, this unitary transformation exhibits conditional dynamics suitable for quantum logic. Setting $\mu_{00} = 0$, $\mu_{10} = \phi_a$, $\mu_{01} = \phi_b$, and defining a parameter $\Delta$ by $\mu_{11} = \phi_a + \phi_b + \Delta$, we find the output state

$$|\psi_{\text{out}}\rangle = |\overline{\alpha}\rangle_a|\overline{\beta}\rangle_b + \overline{\alpha}\overline{\beta}\left[e^{i\Delta} - 1\right]|1\rangle_a|1\rangle_b, \quad (2)$$

where $\overline{\alpha} \equiv \alpha e^{i\phi_a}$ and $\overline{\beta} \equiv \beta e^{i\phi_b}$. This state is clearly entangled for $\Delta \neq 0$. To connect this model to our observations, we examine the reduced density operator for the $a$-field alone and find that in the limit $\beta \to 0$, Eq. (2) leads to $\Phi_a \approx \phi_a - 2m_b \sin(\Delta/2)$. Therefore $\Delta$ may be determined directly from measurements of the initial slope $\partial \Phi_a/\partial m_b$ in a plot of the phase $\Phi_a$ of the probe field versus pump intensity $m_b$.

Note that although the effects of dissipation are neglected in Eqs. (1,2), they could be incorporated via a density matrix corresponding to $|\psi_{\text{out}}\rangle$. However, we shall temporarily set aside such considerations since we are operating with large detunings from atomic resonance in order to approximate purely dispersive interactions. For example, for the measurements of Fig. 3 the amplitude of the probe beam changes by less than 3% in moving from $N = 0$ to $\overline{N} = 1$ intracavity atom.

From the computational point of view, the data of Fig. 3 explicitly demonstrate analog logic (conditional mapping of complex amplitudes) with sub-photon intracavity fields. To make contact with discrete quantum logic, we next consider the relation of our experiment to a *quantum-phase gate* (QPG), for which input Fock states $|1^{\pm}\rangle$ for qubits $(a,b)$ of $\sigma_{\pm}$ polarization are transformed to ouput states with phases specified by the mapping $|1^{\pm}\rangle_a|1^{\pm}\rangle_b \mapsto e^{i\theta_{\pm\pm}}|1^{\pm}\rangle_a|1^{\pm}\rangle_b$ [8]. A sequence of such gates (supplemented by one bit rotations in the $(a,b)$ subspaces) could be combined to serve as a universal element for quantum computation [9]. Our proposed implementation of this gate employs two single-photon pulses $(a,b)$ with frequency separation large compared to the individual bandwidths. These fields would be incident on the cavity mirror M$_2$ of Fig. 1, interact with the atom-cavity system, and then reflect with high efficiency [10]. The basis states $|1^{\pm}\rangle_a|1^{\pm}\rangle_b$ of the truth table for the QPG are associated with $\sigma_{\pm}$ polarizations for the $(a,b)$ fields, which couple to either the weak $g_-$ or strong $g_+$ transition. For $g_- \to 0$, we set $\theta_{--} = 0$ and anticipate that the phase shifts $\theta_{+-}$ and $\theta_{-+}$ will be nothing more than the previously defined phases $(\phi_a, \phi_b)$ for one $\sigma_+$ photon in the $a$ or $b$ mode since, for example, $|1^+\rangle_a|1^-\rangle_b$ should suffer the same phase shift as does $|1^+\rangle_a|0^-\rangle_b$. The dominant nonlinear phase shift should then be $\theta_{++} = \theta_{+-} + \theta_{-+} + \Delta_{++} \simeq \phi_a + \phi_b + \Delta$, with $\Delta \neq 0$ again being the condition for nontrivial dynamics.

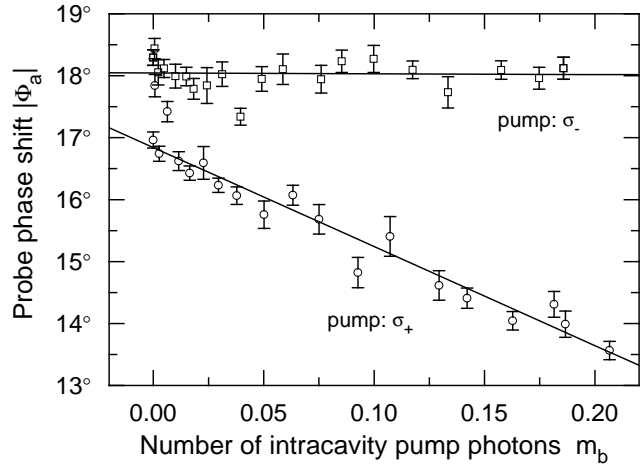

FIG. 4. Dependence of probe phase shift on intensity for two orthogonal polarizations of the pump beam. Pump (probe) detuning is +30 (+20) MHz and $\overline{N} = 0.9$ atoms.

To investigate the truth table for our proposed QPG, we record the dependence of the phase $\Phi_a$ ($\Phi_b$) of the $a$ ($b$) field on the intensity of an injected $b$ ($a$) field of either $\sigma_{\pm}$ polarization, as shown in Fig. 4. Following the discussion of Eq. 2, we extract one-photon phase shifts from initial linear slopes. The straight-line fits shown in Fig. 4 yield $\Delta_{++} \approx \Delta = (16 \pm 3)°$ and $\theta_{+-} - \phi_a = (0.3 \pm 2)° \approx 0$ as anticipated. With the roles of the (a,b) modes interchanged, we can likewise find that $\theta_{-+} \approx \phi_b$. Hence, subject to the validity of our model (1), the experimentally determined parameters for our QPG read

$$\begin{array}{lll}
|1^-\rangle_a|1^-\rangle_b \to & & |1^-\rangle_a|1^-\rangle_b, \\
|1^+\rangle_a|1^-\rangle_b \to & e^{i\phi_a} & |1^+\rangle_a|1^-\rangle_b, \\
|1^-\rangle_a|1^+\rangle_b \to & e^{i\phi_b} & |1^-\rangle_a|1^+\rangle_b, \\
|1^+\rangle_a|1^+\rangle_b \to & e^{i(\phi_a+\phi_b+\Delta)} & |1^+\rangle_a|1^+\rangle_b,
\end{array} \quad (3)$$

where for data as in Fig. 4, $\phi_a \approx (17.5 \pm 1)°$, $\phi_b \approx (12.5 \pm 1)°$, and $\Delta \approx (16 \pm 3)°$.

We believe that this demonstration of *polarization-conditional* phase shifts holds great promise for the implementation of quantum logic with "flying qubits" encoded by the polarization of single-photon pulses. Given the ability to generate a $|1\rangle_a|1\rangle_b$ state of arbitrary polarization, it would then be straightforward to derive states of mutually orthogonal polarization to span the four-dimensional qubit Hilbert space, and hence to measure directly the diagonal elements of the $SU(4)$ transfer matrix (which is a task that cannot be accomplished with only coherent states). Note that single-photon pulses could be generated for this purpose by a variety of techniques and that the optical response of our system to pulses with duration long compared to the inverse cavity damping time $1/\kappa$ should closely reproduce the steady-state behavior investigated here [10]. Furthermore, operation in a regime of strong coupling with $g > \kappa > \gamma$ [7] affords the possibility of yet larger conditional phase shifts for our quantum-phase gate in cavity QED [10].

We wish to stress that the parameter $\Delta$ has model-independent significance as the strength of the dispersive nonlinear interaction between intracavity fields, quoted in degrees per unit of stored energy. Its large measured value represents a unique achievement within the field of nonlinear optics. Our *ansatz* (1) on the other hand may be viewed with some skepticism, for although our assumptions seem reasonable we have not explicitly verified the full transformation (2). We are thus led to consider the question of how to evaluate *operationally* the potential of our system for performing quantum logic, without relying on any particular theoretical model of the appropriate state transformation. From the example provided by Shor's algorithm, it seems reasonable to adopt the observation of coherence and the production of entanglement as *necessary* conditions for calling some candidate device a quantum gate. With these conditions in mind, we turn finally to a brief consideration of experimental strategies for evaluating our laboratory system.

Let us first consider damping of coherences in the output fields by writing their joint density matrix in the generalized form $\rho_{jk} d_{jk}$. Here $\rho_{jk}$ represents a pure-state density matrix in a basis $\{j,k\} = \{0_{a,b}, 1_{a,b}\}$ for Eqs. (1,2) and $\{j,k\} = \{1_{a,b}^-, 1_{a,b}^+\}$ for Eq. (3), and the parameters $d_{jk}$ provide a phenomenological characterization of decoherence. Physical considerations require that $\text{Tr}[\rho_{jk} d_{jk}] = 1$, but dissipative processes could in principle cause complete dephasing of the output density matrix ($d_{j \neq k} \to 0$). Fortunately, with optical fields there exists a straightforward procedure for establishing that this is not the case – heterodyne detection such as implemented in the current work provides signals which are proportional to off-diagonal matrix elements $\rho_{jk} d_{jk}$.

As regards the second criterion, we note that the output state (2) clearly shows entanglement between the pump and probe fields for $\Delta \neq 0$. Hence there must exist a Clauser-Horne-Shimony-Holt (CHSH) inequality [11] violated by correlation measurements on $|\psi_{\text{out}}\rangle$. Following (e.g.) the method of Gisin and Peres [12] we could explicitly formulate the optimal correlation measurement for our particular gate in terms of $\overline{\alpha}$, $\overline{\beta}$, and $\Delta$. Unfortunately the violation must necessarily be of order $|\overline{\alpha}\overline{\beta}(1 - \cos\Delta)|^2 \ll 1$ and therefore quite difficult to detect experimentally. In order to quantify the degree of entanglement which could be generated in our current apparatus we consider the input state $(|1^-\rangle_a + |1^+\rangle_a) \otimes (|1^-\rangle_b + |1^+\rangle_b)/2$, for which the sum of expectation values in the appropriate CHSH inequality is $2\sqrt{1 + \sin^2(\Delta/2)}$. Note that 2 corresponds to the classical upper limit, while the measured conditional phase shift $\Delta \approx 16°$ per photon would generate a value of 2.02 [13]. Although we do not know of any rigorous procedure to compute a "transfer matrix" analogous to (3) for compactly specifying the mapping of input to output states *in the presence of finite dissipation*, the correlation functions appearing in any relevant CHSH inequality can be calculated for arbitrary input fields using Heisenberg equations of motion and the quantum regression theorem. Thus the dependence of entanglement-production on the dissipative parameters $d_{jk}$ introduced above could be investigated in quantitative detail [14].

We acknowledge the contributions of R. J. Thompson, S. Lloyd, A. Ekert and J. Preskill. H. M. holds an NDSEG fellowship. W. L. is supported by DFG. This work is supported by the National Science Foundation (PHY-9014547) and the U. S. Office of Naval Research (N00014-90-J-1058).